\documentclass[conference]{IEEEtran}
\usepackage{pack}

 \usetikzlibrary{external}
 \tikzsetexternalprefix{figures/}

\graphicspath{{figures/}}
\begin{document}

\title{Matched Quantized Min-Sum Decoding of Low-Density Parity-Check Codes}
\author{\IEEEauthorblockN{Emna Ben Yacoub}
	\IEEEauthorblockA{Technical University of Munich\\
		Munich, Germany\\
		Email: emna.ben-yacoub@tum.de}}

\markboth{}{}%

\maketitle

\begin{abstract}
	A quantized message passing decoding algorithm for low-density parity-check codes is presented. The algorithm relies on the min approximation at the check nodes, and on modelling the variable node inbound messages as observations of an extrinsic discrete memoryless channel. The performance of the algorithm is analyzed and compared to quantized min-sum decoding by means of density evolution,  and almost closes the gap with the performance of the sum-product algorithm. A stability analysis is derived, which highlights the role played by degree-$3$ variable nodes in the stability condition. Finite-length simulation results confirm large gains predicted by the asymptotic analysis.
\end{abstract}

\section{Introduction}

	The  deployment of  high throughput communication links \cite{smith_staircase_2012,richardson_design_2018-1} is motivating a revived interest in the design of low-complexity, high-speed channel code decoders. Recently, attention has been devoted to the design and the analysis of iterative decoders where the messages exchanged within the decoder are coarsely quantized. In the context of \ac{LDPC} codes,  message passing algorithms exchanging of binary messages were  presented in \cite{gallagerPhD}. By introducing erasures, the performance of these algorithms is improved  \cite{richardson_capacity}.  Finite-alphabet iterative decoders were also studied, for instance, in \cite{lechner_analysis_2012,planjery2013finite,Bauch,benyacoubTMP,steiner_jlt19}.
While coarse message quantization  reduces the amount of information exchanged within the decoder,  the decoding complexity  can also be reduced  by employing simplified update rules at the \acp{CN}. Examples are the min-sum decoder \cite{Hagenauer96,Fossorier02minsum} and some of its variations (see, e.g., \cite{jones_approximate-min_2003,Lechner06minsum,ryan_channel_codes}), that limit the losses due to the min-approximation at the \acp{CN} by introducing simple corrections.

In this paper, we  analyze and  design quantized min-sum decoders \cite{kschischang_minsum}.  At the \acp{CN}, we use the standard min-approximation rule. In contrast to the \ac{QMS} algorithm \cite{kschischang_minsum}, the \ac{VN} decoder converts all incoming messages to \acp{LLR} by modeling the extrinsic channel as a \ac{DMC}, extending the approach introduced  for binary message passing decoding in \cite{lechner_analysis_2012} to the case where messages are represented by $b$ bits. The transition probabilities of the extrinsic \acp{DMC} are derived via \ac{DE} analysis, which is developed for unstructured irregular \ac{LDPC} ensembles. Because the \ac{VN} inbound messages are matched to the reliability of the underlying extrinsic \ac{DMC}, we refer to the proposed algorithm as \ac{MQMS} decoding. A stability analysis is derived. As observed for the  quantized message passing decoders of \cite{lechner_analysis_2012,benyacoubTMP,steiner_jlt19}, the fraction of edges connected to degree-$3$ \acp{VN} plays an important role in the stability condition of the proposed decoding algorithm. The \ac{DE} analysis shows how the proposed \ac{MQMS} decoding algorithm  improves the decoding thresholds of the \ac{QMS} decoder \cite{kschischang_minsum},  closing the gap  to the performance achieved by the \ac{SPA}. The results are confirmed by finite-length simulations.

\section{Preliminaries}\label{sec:prelimin}

\subsection{LDPC Codes}
	\ac{LDPC} codes are binary linear block codes defined by an $ m \times n $ sparse parity-check matrix $\bm{H}$. The code dimension is $k \geq n - m$. The Tanner graph of an \ac{LDPC} code is a bipartite graph $G = (\cV \cup \cC, \cE)$ consisting of $n$ \acp{VN} and $m$ \acp{CN}. The set $\cE$ of edges contains the elements
$e_{ij}$, where $e_{ij}$ is an edge between \ac{VN} $\vn_j\in\cV$ and \ac{CN} $\cn_i\in\cC$. Note that $e_{ij} \in \cE$ if and only if the parity-check matrix element $h_{ij}$ is equal to $1$. The sets $\cN(\vn_j)$ and $\cN(\cn_i)$ denote the neighbors of \ac{VN} $\vn_j$ and \ac{CN} $\cn_i$, respectively. 
The degree  of a \ac{VN} $\vn_j$ (\ac{CN} $\cn_i$) is  the cardinality of the set $\cN(\vn_j)$ ($\cN(\cn_i)$). The edge-oriented degree distribution polynomials of an \ac{LDPC} code graph are
$\lambda(x)= \sum_{i}\lambda_{i}x^{i-1}$ and  $\rho(x)= \sum_{i}\rho_{i}x^{i-1}$ 
where $ \lambda_{i} $ and $ \rho_{i} $ are, respectively, the fraction of edges incident to \acp{VN} and \acp{CN} with degree $ i $.  An unstructured irregular \ac{LDPC} code ensemble $ \ensC $ is the set of all \ac{LDPC} codes with block length $ n $ defined by a bipartite graph with degree distributions $ \lambda\left( x\right) $ and $ \rho\left( x\right) $. 

	\subsection{Channel Model}
We consider the \ac{biAWGN} channel with input alphabet $ \cX  = \{-1, +1 \}$. The channel output is $ Y = X + N $, where $ N $ is  Gaussian \ac{RV} with zero mean and variance $\sigma^{2} $.  The channel \ac{SNR} is  $ E_{\tb}/N_{0} $, where $ E_{\tb} $ is the energy per information bit and $ N_{0} $ is the single-sided noise power spectral density.

\subsection{Extrinsic Channels} \label{sec:channel}
Consider   a  binary-input $M$-ary output \ac{DMC} with input alphabet $\cX = \{-1,+1\}$ and output alphabet $\cZ=\{-\frac{M-1}{2},-\frac{M-3}{2},\ldots,0,\ldots,\frac{M-1}{2} \}$, where $M=2^b-1$ and $b$ is a positive integer. For a generic channel output $z$,  \acp{LLR} can be obtained as 
\begin{equation}
\llr(z) = \ln\left[ \frac{P_{Z|X}(z|+1)}{P_{Z|X}(z|-1)}\right].
\end{equation}
If the channel satisfies the symmetry constraint  \linebreak $P_{Z|X}(-z|+1) =   P_{Z|X}(z|-1) $ for all $ z \in \cZ$, we have
\begin{equation}\label{eq:LLRdecomposition}
\llr(z)=
\sign(z) \D_{\abs{z}} 
\end{equation}
where $\forall a \in \cZ, a > 0$
\begin{equation} \label{eq:Dval}
\D_a:=
\ln\left[ \frac{P_{Z|X}(a|+1)}{P_{Z|X}(-a|+1)}\right]
\end{equation}
and where by convention the $\sign(x)$ function takes value $0$ for $x=0$.
We refer to  $\D_{\abs{z}} $ as the \emph{reliability} of $z$.
The decomposition  \eqref{eq:LLRdecomposition} will be instrumental to the development of a message-passing decoding algorithm for \ac{LDPC} codes. In particular, we will focus on  a decoding algorithm that exchanges quantized messages. In this case, a message sent from a \ac{CN} to a \ac{VN} can be modeled as the observation of the \ac{RV} $X$ after transmission over a binary-input $M$-ary output discrete memoryless extrinsic channel \cite[Fig.~3]{ashikhmin_EXIT}, where $M$ is the number of message quantization levels. While the transition probabilities of the extrinsic channel are in general unknown,  accurate estimates can be obtained via \ac{DE} analysis, as suggested in \cite{lechner_analysis_2012}. This observation will be used to derive the decoding algorithm presented in Section \ref{sec:algorithms}.

\subsection{Quantization}\label{sec:prelimin:quant}

Throughout the paper, we  consider uniform quantization of the messages. We denote by $\qf: \mathbb{R} \to \cM$  the quantization function of the exchanged messages, where the quantized message alphabet is $ \cM=\{ -\Sm\Delta,-(\Sm-1)\Delta, \ldots, \Sm \Delta \} $. The function $\qf$ is a $b$-bit uniform quantizer with step size $\Delta$ and $2^b -1$ quantization levels. Formally, we have
\begin{align}\label{eq:f}
\qf(x) &:= \sign(x) \Delta \cdot \min \left\{ \left\lfloor  \frac{\abs{x}}{\Delta} + \frac{1}{2} \right\rfloor,  \Sm \right\}
\end{align}
where $\Sm =2^{b-1} -1$.

For the channel output, we  consider two cases: unquantized  channel outputs and quantized  channel outputs.
For the latter case, the \ac{biAWGN} channel output is quantized using a $\bch$-bit uniform quantizer with step size $\Deltach$, where $\bch$ and $\Deltach$ may in general differ from the corresponding parameters for the message quantization. The quantized channel output alphabet is  $\cM_{0}= \{-\Sch\Deltach,-(\Sch-1)\Deltach, \ldots, \Sch\Deltach  \}$ with $\Sch=2^{b_{0}-1} -1$, and the quantized version of $y$ is denoted as $\mch$.

\section{Matched Quantized Min-Sum Decoding}\label{sec:algorithms}

We denote by  $ \mcv{\ell}  $  the message sent from \ac{CN} $ \cn $ to its neighboring \ac{VN} $ \vn $. Similarly, $ \mvc{\ell}  $ is the message sent from \ac{VN} $ \vn $ to \ac{CN} $ \cn $ at the $\ell$-th iteration. 

\subsection{Unquantized Channel Output}
Each \ac{VN} computes the \ac{LLR} of the corresponding channel output
\begin{equation} \label{eq:channelLLRsoft}
\lch(y)=\frac{2}{\sigma^2}y.
\end{equation}
Then the \ac{VN} passes a $b$-bit quantized value to its neighboring \acp{CN}. Thus, $\forall \cn \in \cNv$ we have
\begin{equation}
\mvc{0}= \qf( \lch(y))
\end{equation}
where $\qf$ is defined in \eqref{eq:f} and we choose $\Delta$ to minimize the iterative decoding threshold.

The min update rule is performed at the \acp{CN}. We have
\begin{equation} \label{eq:CNminsum}
\mcv{\ell} = \min\limits_{\vn' \in \cN(\cn) \setminus \vn} \abs{\mvcex{\ell-1}} \prod\limits_{\vn' \in \cN(\cn) \setminus \vn} \sign\left(\mvcex{\ell-1}\right).
\end{equation}
At the $\ell$-th iteration, each \ac{VN} converts its channel message and the incoming \ac{CN} messages to \acp{LLR}.  The sum of these \acp{LLR} is then quantized into a $b$-bit message. Formally, we have
\begin{equation}\label{eq:vnQDSC}
\mvc{\ell}=\qf \left(\lch(y) + \sum\limits_{\cn' \in \cN(\vn) \setminus \cn }\lex\left(\mcvex{\ell}\right) \right)
\end{equation}
where 
\begin{equation} \label{eq:llrmessages}
\lex\left(\mcvex{\ell}\right)  :=
\sign\left(\mcvex{\ell}\right) \Dex{\ell}_{\abs{\mcvex{\ell}}}.
\end{equation}
The final hard decision at  each \ac{VN} is
\begin{equation}\label{eq:appQDSC}
\hat{x}_{\vn}^{(\ell) } = \sign \left(\lch(y) + \sum\limits_{\cn' \in \cN(\vn) }\lex\left(\mcvex{\ell}\right) \right)
\end{equation}

\begin{figure*}[t]
	\centering
	\begin{equation} \label{eq:cnprob}
	\q{i}{\ell}=
	\begin{cases}
	\frac{1}{2}\left[ \rho\left(\A{i}{\ell-1} +\B{i}{\ell-1} \right) +  \rho\left(\A{i}{\ell-1} -\B{i}{\ell-1} \right) -\rho\left(\A{i+1}{\ell-1} +\B{i+1}{\ell-1} \right) -  \rho\left(\A{i+1}{\ell-1} -\B{i+1}{\ell-1} \right)\right] & i > 0 \\
	1-\rho\left(1-\p{0}{\ell-1}\right) & i= 0 \\
	\frac{1}{2}\left[ \rho\left(\A{-i}{\ell-1} +\B{-i}{\ell-1} \right) -  \rho\left(\A{-i}{\ell-1} -\B{-i}{\ell-1} \right) -\rho\left(\A{-i+1}{\ell-1} +\B{-i+1}{\ell-1} \right) + \rho\left(\A{-i+1}{\ell-1}-\B{-i+1}{\ell-1} \right)\right] & i < 0
	\end{cases}
	\end{equation}\vspace{-3mm}
	\begin{align}\label{eq:vnprobsoft}
	\p{i}{\ell} = \begin{cases}
	\sum\limits_{d} \lambda_d \sum\limits_{\lin}\Pr\left\{\Lin^{(\ell)}=\lin\right\}	Q\left(\frac{(\Sm-\frac{1}{2})\Delta+\lin+\mu_\tch}{\sigma_\tch}\right)  & i=-\Sm\\
	\sum\limits_{d} \lambda_d \sum\limits_{\lin}\Pr\left\{\Lin^{(\ell)}=\lin\right\}	Q\left(\frac{(\Sm-\frac{1}{2})\Delta-\lin-\mu_\tch}{\sigma_\tch}\right) & i=\Sm\\
	\sum\limits_{d} \lambda_d \sum\limits_{\lin}\Pr\left\{\Lin^{(\ell)}=\lin\right\}	\left[Q\left(\frac{(i-\frac{1}{2})\Delta-\lin-\mu_\tch}{\sigma_\tch}\right)-Q\left(\frac{(i+\frac{1}{2})\Delta-\lin-\mu_\tch}{\sigma_\tch}\right)\right] & \text{otherwise}	
	\end{cases}
	\end{align}
	\vspace{-2mm}
	\hrule
\end{figure*}

Note that the reliability of $\mcvex{\ell}$ depends on the iteration number and is in general unknown. In fact, the transition probabilities of the underlying extrinsic \acp{DMC} are not known. As proposed in \cite{lechner_analysis_2012}, their values can be estimated via Monte Carlo simulations, or via \ac{DE} analysis. The latter approach provides accurate results for moderate to large block lengths, as shown in \cite{lechner_analysis_2012,benyacoubTMP}. We hence follow this direction and use the \ac{DE} presented in Section \ref{sec:densityevolution} to estimate the message reliability at each iteration. For the special case of $b=2$, we will obtain the \ac{TMP} decoder introduced in \cite{benyacoubTMP}.

\subsection{Quantized Channel Output}

If the channel output is quantized as described in Section \ref{sec:prelimin:quant}, we replace $\lch(y)$ in \eqref{eq:vnQDSC} and \eqref{eq:appQDSC} by $\lch (\mch)=
\sign(\mch) D_{\abs{\mch}}$.  We choose $\Delta$ and $\Deltach$ to minimize the decoding threshold.
As mentioned in Sec.~\ref{sec:channel}, the decoder's communication channel can be modeled as a binary-input $\vert \cM_0\vert$-ary output \ac{DMC} that satisfies the symmetry condition. The value of  $\D_{\abs{\mch}}$  can then  be computed from \eqref{eq:Dval} by using the transition probabilities of the quantized communication channel.

\section{Density Evolution Analysis}\label{sec:densityevolution}
We provide a \ac{DE} analysis of the \QD  algorithm for unstructured \ac{LDPC} code ensembles. Due to symmetry, we may assume that the all-zeros codeword is transmitted. Let $\Mvc{\ell}$  be the \ac{RV} associated to \ac{VN} to \ac{CN} messages at the $\ell$-th iteration. Similarly, $\Mcv{\ell}$ represents the \ac{RV} associated to \ac{CN} to \ac{VN} messages. We denote by $\p{i}{\ell}$ the probability that $\Mvc{\ell}$  takes the value $\Delta i$, with $i\in \{-\Sm, -(\Sm-1),\ldots,\Sm\}$. Similarly, we denote by $\q{i}{\ell}$ the probability that $\Mcv{\ell}$ takes the value $\Delta i$. In the following, $\ell_{\tmax}$ denotes the maximum number of iterations. 
In the limit of $n \to \infty$, the evolution of the message distributions can be tracked as follows.
\begin{enumerate}
	\item \textbf{Initialization.}  Conditioned on $ X = +1$, the channel \acp{LLR} are Gaussian \acp{RV} with mean  $\mu_{\tch}= 4RE_{\tb}/N_{0} $ and variance $ \sigma_{\tch}^{2}= 2 \mu_{\tch} $. Therefore, we have
	\begin{align}
	\p{i}{0} = \begin{cases}
	Q\left(\frac{(\Sm-\frac{1}{2})\Delta+\mu_\tch}{\sigma_\tch}\right)  &i=-\Sm\\
	Q\left(\frac{(\Sm-\frac{1}{2})\Delta-\mu_\tch}{\sigma_\tch}\right) &i=\Sm\\
	Q\left(\frac{(i-\frac{1}{2})\Delta-\mu_\tch}{\sigma_\tch}\right)-Q\left(\frac{(i+\frac{1}{2})\Delta-\mu_\tch}{\sigma_\tch}\right) & \text{otherwise}	
	\end{cases}
	\end{align}
	while if the channel output is quantized we have 
	\begin{equation}
	\p{i}{0}= \sum\limits_{\mch:\qf(\lch(\mch))=\Delta i} P_{\Mch|X}(\mch|+1).
	\end{equation}
	\item \textbf{For} $ \ell =1, 2, \ldots, \ell_{\tmax} $
	
	\smallskip	
	
	\textbf{Check to variable update.} For all $j \in \{1,\ldots,2^{b-1} \}$, we  define $\A{j}{\ell}$ and $\B{j}{\ell}$ as
	\begin{align}
	\A{j}{\ell} &:= \Pr\left\{\Mvc{\ell} \geq \Delta j \right\}\\ 
	\B{j}{\ell} &:= \Pr\left\{\Mvc{\ell} \leq -\Delta j \right\}.
	\end{align}
	The probabilities $\q{i}{\ell}$ can be computed as shown in \eqref{eq:cnprob}.
	
	\smallskip
	\textbf{Variable to check update.} For $i \in \{ -\Sm, -(\Sm-1),\ldots,\Sm \} $,  $\p{i}{\ell}$ can be computed from \eqref{eq:vnprobsoft} for the unquantized channel output, while for the quantized one we have
	\begin{equation}\label{eq:vnprobquant}
	\begin{aligned}
	\p{i}{\ell} =& \sum\limits_{d} \lambda_d \sum\limits_{\mch}P_{\Mch|X}(\mch|+1) \times \\&\sum\limits_{\lin: \qf(\lch(\mch)+\lin)=\Delta i}\Pr\left\{\Lin^{(\ell)}=\lin\right\}
	\end{aligned}
	\end{equation}
	where $\Lin^{(\ell)}$ is a \ac{RV} associated to the sum of the \acp{LLR} of the $d-1$ incoming \ac{CN} messages at the $\ell$-th iteration. We have
	\begin{equation}\label{eq:Linprob}
	\Pr\left\{\Lin^{(\ell)}=\lin\right\} = \sum\limits_{\bm{v}} \binom{d-1}{v_{-\Sm},\ldots,v_{\Sm}}  \prod\limits_{i=-\Sm}^{\Sm} \left(\q{i}{\ell}\right)^{v_i}
	\end{equation}
	where the sum is over all integer vectors $\bm{v}$ for which
	\begin{equation}
	\sum\limits_{i=-\Sm}^{\Sm} v_i=d-1 \quad \text{and} \quad  \sum\limits_{i=1}^{\Sm} (v_i-v_{-i})  \Dex{\ell}_{\Delta i}=\lin
	\end{equation}
	where
	$\Dex{\ell}_{\Delta i} := \ln \left( \q{i}{\ell}/\q{-i}{\ell}\right)$.
	Note that the vector entry $v_i$ represents the number of incoming \ac{CN} messages with value $\Delta i$.
\end{enumerate}

\smallskip

The ensemble iterative decoding threshold $ (E_{\tb}/N_{0})^\star $ is defined as the minimum $ E_{\tb}/N_{0} $ for which $ \lim\limits_{ \ell \to \infty}	P_e^{(\ell)} = 0$ as $n \to \infty$, where 
\begin{equation}
P_e^{(\ell)}=\sum\limits_{i=-\Sm}^{0}\p{i}{\ell}. 
\end{equation}

\section{Stability Analysis}\label{sec:stability}
We define 
\begin{align}
\vp{\ell} := \left[ \p{-\Sm}{\ell}, \ldots,  \p{\Sm-1}{\ell} \right]^{T} \quad \!\text{and} \!\quad  \vq{\ell} := \left[ \q{-\Sm}{\ell} ,\ldots, \q{\Sm-1}{\ell} \right]^{T}\!\!.\\[-8mm]
\end{align}
We should determine the evolution of $\vp{\ell} $ over one iteration in proximity of the fixed point $ \pst= \bm{0} $.\footnote{We  assume  that the minimum \ac{VN} degree is at least $2$.} Note that, as $\vp{\ell} \to \bm{0} $,  $ \vq{\ell} \to \bm{0} $. Thus, for the inbound \ac{VN} extrinsic channel, we have $\D_{i \Delta}^{(\ell)} \to +\infty$ for $i=S$ while $\D_{i \Delta}^{(\ell)} \to 0$ for $i<S$.

\medskip

As $\vq{\ell} \to \bm{0}$, we have 
\begin{align}\label{eq:p1stab}
\begin{split}
\p{-\Sm}{\ell}=\sum\limits_{d}\lambda_{d} &\left[ \Pr\left\{\Sin{\ell}=0\right\} \Pr\left\{L_{\tch} \leq -(\Sm-0.5)\Delta \right\} \right. \\ &\left.+\Pr\left\{\Sin{\ell} \leq -1 \right\} \right]
\end{split}
\end{align}
while for $i \in \{ -\Sm+1, \ldots, \Sm-1 \}$
\begin{equation}\label{eq:p2stab}
\p{i}{\ell}\!\!=\!\!\sum\limits_{d}\!\lambda_{d} \Pr\left\{\Sin{\ell}\!=0\right\} \Pr\left\{(i-0.5)\Delta \leq L_{\tch} \leq (i+0.5)\Delta \right\}  
\end{equation}
where $\Sin{\ell} $ is a \ac{RV} representing the difference between the number of incoming messages equal to $ \Sm$ and the number of incoming messages equal to $-\Sm$ to a \ac{VN} of degree $d$. We have
\begin{equation}\label{eq:Sinprob}
\Pr\left\{\Sin{\ell}=\sins\right\} = \sum\limits_{\bm{v}} \binom{d-1}{v_{-\Sm},\ldots,v_{\Sm}}  \prod\limits_{i=-\Sm}^{\Sm} \left(\q{i}{\ell}\right)^{v_i}
\end{equation}
where the sum is over all integer vectors $\bm{v}$ for which
\[
\sum\limits_{i=-\Sm}^{\Sm} v_i=d-1, \quad   (v_{\Sm}-v_{-\Sm}) =\sins.
\]
We further have
\begin{equation}\label{eq:diffS1}
\lim\limits_{\vq{\ell} \to \bm{0}} \frac{\partial \Pr\left\{ \Sin{\ell}=\sins \right\}}{\partial \q{-\Sm }{\ell}   } = \begin{cases}
d- 1 &  \sins =d-3 \\
-(d-1) & \sins =d-1 \\
0 & \text{ otherwise}
\end{cases}
\end{equation}
while, for $i \in \{ -\Sm+1, \ldots, \Sm-1 \}$,
\begin{equation}\label{eq:diffS2}
\lim\limits_{\vq{\ell} \to \bm{0}} \frac{\partial \Pr\left\{ \Sin{\ell}=\sins \right\}}{\partial \q{i }{\ell}   } = \begin{cases}
d- 1 &  \sins =d-2 \\
-(d-1) & \sins =d-1 \\
0 & \text{ otherwise}.
\end{cases}
\end{equation}
From  \eqref{eq:cnprob}, for $i,j \in \left\{ -\Sm, \ldots, \Sm-1 \right\}$
\begin{equation}\label{eq:diffpq}
\lim\limits_{\vp{\ell-1} \to \bm{0}} \frac{ \partial \q{i}{\ell}}{\partial  \p{j}{\ell-1} }= \begin{cases}
\rho'(1) & i=j \\
0 & \text{ otherwise}.
\end{cases}
\end{equation}
From \eqref{eq:p1stab}, \eqref{eq:Sinprob}, \eqref{eq:diffS1}, \eqref{eq:diffS2}, \eqref{eq:diffpq}, we obtain
\begin{equation}\label{eq:stab1}
\lim\limits_{\vp{\ell-1} \to \bm{0}} \frac{\partial \p{-\Sm}{\ell}}{\partial  \p{-\Sm}{\ell-1} }= \rho'(1) \left(2 \lambda_3 \alpha + \lambda_2\right)
\end{equation}
for $i,j \in \left\{ -\Sm+1, \ldots, \Sm -1 \right\}$
\begin{equation}\label{eq:stab2}
\lim\limits_{\vp{\ell-1} \to \bm{0}} \frac{\partial \p{-\Sm}{\ell}}{\partial  \p{i}{\ell-1} }=   \lambda_2 \alpha \rho'(1)
\end{equation}
\begin{equation}\label{eq:stab3}
\lim\limits_{\vp{\ell-1} \to \bm{0}} \frac{\partial \p{i}{\ell}}{\partial  \p{-\Sm}{\ell-1} }= 2\rho'(1)  \lambda_3 \gamma_i
\end{equation}
\begin{equation}\label{eq:stab4}
\lim\limits_{\vp{\ell-1} \to \bm{0}} \frac{\partial \p{i}{\ell}}{\partial  \p{j}{\ell-1} }= \rho'(1)  \lambda_2 \gamma_i 
\end{equation}
where 
\begin{align}
\alpha &=Q\left( \frac{(\Sm-\frac{1}{2})\Delta+\mu_{\tch}}{\sigma_{\tch}}\right)\\
\gamma_i&=Q\left( \frac{(i-0.5)\Delta-\mu_{\tch}}{\sigma_{\tch}}\right) - Q\left( \frac{(i+0.5)\Delta-\mu_{\tch}}{\sigma_{\tch}}\right)
\end{align}
if the channel output is unquantized while 
\begin{align}
\alpha &=\sum\limits_{\mch:\qf(\lch(\mch))=-\Sm \Delta } P_{\Mch|X}(\mch|+1)\\
\gamma_i&= \sum\limits_{\mch:\qf(\lch(\mch))=\Delta i} P_{\Mch|X}(\mch|+1)
\end{align}
if the channel output is quantized. The first order Taylor expansions via \eqref{eq:stab1}, \eqref{eq:stab2}, \eqref{eq:stab3}, \eqref{eq:stab4} yield
\begin{equation}
\vp{\ell}= 
\bm{A}\cdot \vp{\ell-1}
\end{equation} 
where for $i,j \in  \left\{ -\Sm, \ldots, \Sm-1 \right\}$
\begin{equation}
a_{i,j}= 	\lim\limits_{\vp{\ell-1} \to \bm{0}} \frac{\partial \p{i}{\ell}}{\partial  \p{j}{\ell-1} }.
\end{equation}
Let $ r $ be the spectral radius of $ \bm{A} $. The stability condition is fulfilled if and only if $  r < 1 $.

\def\arraystretch{1.2}
\begin{table}[t]
	\centering
	\caption{Decoding thresholds $(E_{b}/N_0)^\star  [\si{dB}]$ of \QD for quantized and unquantized channel output and for \ac{QMS}.}
	\begin{tabular}{c|c|c|c|c|c}
		\hline\hline  
		\multirow{2}{*}{$(\dv,\dc)$} & \multirow{2}{*}{$b$}              & \QD    & \multirow{2}{*}{$\bch$}   & \QD   & \multirow{2}{*}{QMS}       \\ 
		&            &  (unquant. channel)              &  & (quant. channel)  &      \\
		\hline
		&\multirow{3}{*}{2}& \multirow{3}{*}{1.85}  & 2       & 2.39      & 2.66             \\ 
		&                  &                        & 3       & 1.9       & 2.66             \\ 
		&                  &                        & 4       & 1.86      & 2.58           \\ 
		\cline{2-6}  
		&\multirow{2}{*}{3}&\multirow{2}{*}{1.32}	& 3       & 1.45              & 1.8                \\
		(3,6)   &                  &                        & 4       & 1.34              & 1.8                \\
		\cline{2-6}  
		& \multirow{2}{*}{4} & \multirow{2}{*}{1.21}	& 3       & 1.35              & 1.72              \\
		&                    &                          &  4      & 1.24              & 1.65               \\
		\cline{2-6}  
		& 5	                 &1.18                      & 5       & 1.19              & 1.62              \\
		\hline
		& \multirow{3}{*}{2}&\multirow{3}{*}{2.11}   	& 2       & 2.71              & 2.78                \\
		&                   & 	                        & 3       & 2.22              & 2.4                \\
		&                   &                           & 4       & 2.11              & 2.43              \\
		\cline{2-6}  
		& \multirow{2}{*}{3}  & \multirow{2}{*}{1.73} & 3     & 1.85              & 2.17              \\
		(4,8)   &                     &                       & 4     & 1.76              & 2.12               \\
		\cline{2-6}  
		& \multirow{2}{*}{4}  & \multirow{2}{*}{1.65} 	& 3     & 1.77              & 2.14              \\
		&                     &                         & 4     & 1.68              & 2.08               \\
		\cline{2-6}  
		& 5                   & 1.63                      & 5 &1.64 & 2.06              \\
		\hline\hline 
	\end{tabular}
	\label{tab:threhsoldsreg} 
\end{table}

\begin{remark}
	By a close inspection of \eqref{eq:stab1} and \eqref{eq:stab3}, we see that the fraction of edges connected to degree-$3$ \acp{VN} plays an important role in the stability condition of the proposed decoding algorithm. This was already noted for the  quantized decoders of \cite{lechner_analysis_2012,benyacoubTMP,steiner_jlt19},  and in the analysis of saturated belief propagation decoding of \cite{Kudekar_saturated}. The result hence shows that one should more strongly limit
	the use of degree-$2$ and degree-$3$ \acp{VN} for  unstructured \ac{LDPC} ensembles, with respect to the case of unquantized, non-saturated belief propagation decoding. From a practical viewpoint, it might be worth relaxing the definition of decoding threshold by adopting a suitably-low target decoding error probability, especially for codes  tailored for moderate error rates.
\end{remark}

\section{Numerical Results}\label{sec:results}
\begin{figure}[t]
	\centering
	\footnotesize
	\includegraphics{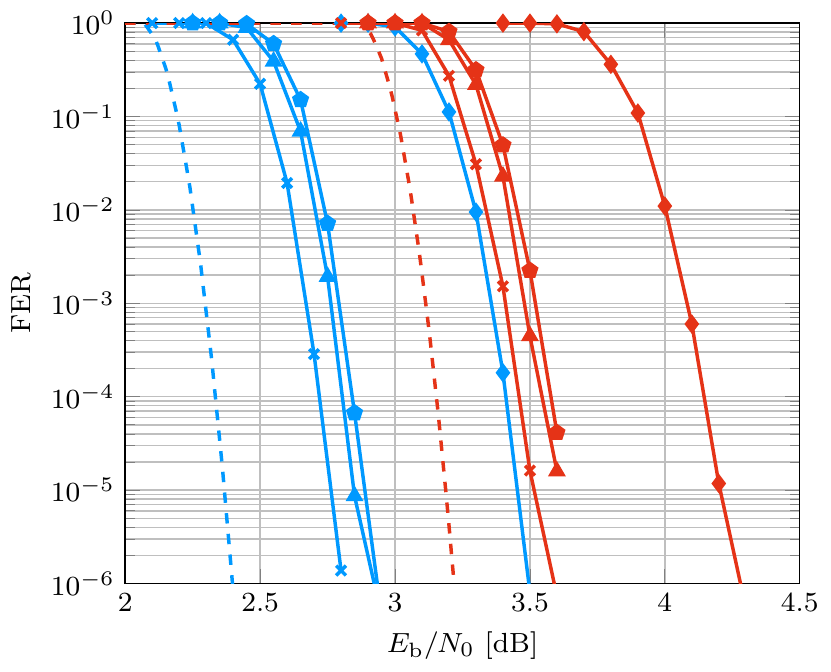}
	\caption{FER versus $E_\tb/N_0$ [\si{dB}] for unquantized \ac{SPA} (\protect\tikz[baseline=-0.5ex]{\protect\draw[line width=1,color=TUMBeamerBlue]  plot[mark=x, mark options={solid, scale=0.8}] (0,0) -- (0.4,0)} ,\protect\tikz[baseline=-0.5ex]{\protect\draw[line width=1,color=TUMBeamerRed]  plot[mark=x, mark options={solid, scale=0.8}] (0,0) -- (0.4,0)} ),  \ac{MQMS} for unquantized  (\protect\tikz[baseline=-0.5ex]{\protect\draw[line width=1,color=TUMBeamerBlue]  plot[mark=triangle*, mark options={solid, scale=0.8}] (0,0) -- (0.4,0)} ,\protect\tikz[baseline=-0.5ex]{\protect\draw[line width=1,color=TUMBeamerRed]  plot[mark=triangle*, mark options={solid, scale=0.8}] (0,0) -- (0.4,0)}) and quantized (\protect\tikz[baseline=-0.5ex]{\protect\draw[line width=1,color=TUMBeamerBlue]  plot[mark=pentagon*, mark options={solid, scale=0.8}] (0,0) -- (0.4,0)} ,\protect\tikz[baseline=-0.5ex]{\protect\draw[line width=1,color=TUMBeamerRed]  plot[mark=pentagon*, mark options={solid, scale=0.8}] (0,0) -- (0.4,0)}) channel output, \ac{QMS} (\protect\tikz[baseline=-0.5ex]{\protect\draw[line width=1,color=TUMBeamerBlue]  plot[mark=diamond*, mark options={solid, scale=0.8}] (0,0) -- (0.4,0)} ,\protect\tikz[baseline=-0.5ex]{\protect\draw[line width=1,color=TUMBeamerRed]  plot[mark=diamond*, mark options={solid, scale=0.8}] (0,0) -- (0.4,0)}), RCU bound (\protect\tikz[baseline=-0.5ex]{\protect\draw[dashed,line width=1,TUMBeamerBlue] (0,0) -- (.4,0)} ,\protect\tikz[baseline=-0.5ex]{\protect\draw[dashed,line width=1,TUMBeamerRed] (0,0) -- (.4,0)}) for $R=4/5$ (\protect\tikz[baseline=-0.5ex]{\protect\draw[line width=1,TUMBeamerBlue] (0,0) -- (.4,0)}), $R=7/8$ (\protect\tikz[baseline=-0.5ex]{\protect\draw[line width=1,TUMBeamerRed] (0,0) -- (.4,0)}).}
	\label{fig:results_finitelength}
\end{figure}

A first set of results deals with the asymptotic performance of \ac{MQMS} decoding. Table \ref{tab:threhsoldsreg} reports a comparison between the iterative decoding thresholds of \QD for both quantized and unquantized channel outputs and \ac{QMS} \cite{kschischang_minsum} for $(\dv,\dc)$ regular \ac{LDPC} ensembles and different values of $b$ and $\bch$. \ac{MQMS} decoding largely outperforms \ac{QMS}, with gains up to $\SI{0.7}{dB}$. Remarkably, for $b=\bch=5$ the \QD thresholds are within $\SI{0.1}{dB}$ of the unquantized belief propagation thresholds (which are at $(E_{b}/N_0)^\star \approx \SI{1.1}{dB}$ for the regular $(3,6)$ ensemble, and at $(E_{b}/N_0)^\star \approx \SI{1.58}{dB}$ for the regular $(4,8)$ ensemble).
Based on the \ac{DE} analysis of Section \ref{sec:densityevolution}, we designed a set of optimized irregular ensembles with various rates. For the design, we assumed a \QD decoder with $b=4$ and unquantized channel output. We set the maximum \ac{VN} degree to $\mv=20$. Due to space limitations, the optimized degree distributions, obtained via differential evolution are provided in \cite{BenYacoubMQMSarxiv}.
We next considered the performance for rate  $4/5$ and $7/8$ codes, designed for  \QD decoder and unquantized channel outputs, where we set $b=4, \mv=15, \ell_{\tmax} =30$. The codes have a block length $ n= 20000$ bits and their graphs were designed via the \ac{PEG} algorithm \cite{Hu_peg}. The simulation results are shown in Fig.~\ref{fig:results_finitelength}  in terms of \ac{FER} versus $E_{\tb}/N_0$. As a reference, we  provide the simulation results of the optimized codes for \QD under unquantized \ac{BP} decoding, \QD for both $4$ bit quantized and unquantized channel output and \ac{QMS} with $b =\bch=4$, as well as the random coding union bound (RCU) of \cite{polyanskiy_channel_2010}. Observe that the \ac{MQMS} algorithm outperforms the \ac{QMS} decoder although they both use the same \ac{CN} update rule. Admittedly, the \ac{VN} update rule of \QD is more complex than the one of the plain \ac{QMS} decoder: An open question is whether the \ac{VN} update rule in \eqref{eq:vnQDSC} can be efficiently implemented in approximate form (e.g., via look-up tables) without compromising the performance of the \ac{MQMS} algorithm.

\section{Conclusion}\label{sec:conclusions}
A quantized message passing decoding algorithm for \ac{LDPC} codes was presented. The algorithm relies on the min approximation at the check nodes, and on modelling the variable node inbound messages as observations of an extrinsic channel output. The algorithm was analyzed via density evolution. Degree-$3$ variable nodes play an important  role in the stability condition. The algorithm shows remarkable gains over quantized min-sum decoding, almost closing the gap with the performance of the \ac{SPA}.


\def\arraystretch{1.2}
\begin{table*}[h]
	\centering
	\caption{Thresholds of optimized degree distributions for the \QD decoder for unquantized channel with $b=4$ and quantized channel with $b=\bch=4$.}
	\begin{tabular}{c|c|c|c|c|c } \hline\hline
		\multirow{2}{*}{$R$}&\multirow{2}{*}{$\lambda(x)$}&\multirow{2}{*}{$\rho(x)$}&  \multirow{2}{*}{$\left(E_{\text{b}}/N_0\right)^\star[\si{dB}]$} &$\left(E_{\text{b}}/N_0\right)^\star[\si{dB}]$& \multirow{2}{*}{$\left(E_{\text{b}}/N_0\right)_{\text{Sh}}[\si{dB}]$} \\
		& & & & $\bch=4$&  \\
		\hline
		$2/3$ &  $0.0317 x+ 0.489 x^2 +0.0374 x^{9}+0.4419 x^{19}$ & $0.328 x^{13}+0.672 x^{14} $&$1.47$&$1.5  $ & $1.06$ \\
		\hline
		$3/4$ &  $ 0.0313 x+0.463 x^2+0.0058x^9+0.4999 x^{19}$& $0.5336 x^{19}+0.4664 x^{20}  $ & $1.96 $ 
		&$2$ & $ 1.62$\\
		\hline
		$4/5$ &  $ 0.4961 x^2+0.0051 x^9+0.4988 x^{19} $& $ 0.7907x^{25}+0.2093 x^{26}$& $ 2.34$&$ 2.37$ & $2.04$\\
		\hline
		$5/6$ &  $ 0.0205x+0.4646 x^2+0.0534 x^{9}+0.4616 x^{19}$& $0.9926 x^{30} +0.0074 x^{31}  $  & $2.63 $& $2.66$ & $ 2.36$\\
		\hline
		$7/8$ &  $0.4789 x^2+0.0021x^4+0.032 x^9+0.487 x^{19} $& $0.3752 x^{41}+0.6248 x^{42}  $  & $3.08 $ &$3.11$ &$ 2.85$ \\ 
		\hline
		$9/10$ &  $ 0.4442 x^2+0.0403 x^3+0.0025 x^9+0.513 x^{19}$& $ 0.6604 x^{53}+0.3396 x^{54} $ & $ 3.42$ &$3.44$ & $3.2 $\\ \hline\hline
	\end{tabular}
	\label{tab:thresholdsQDSC}
\end{table*}

\section*{Acknowledgment}
The author is grateful to Dr. Gianluigi Liva for many  fruitful discussions and valuable comments.
\bibliographystyle{IEEEtran}
\bibliography{IEEEabrv,confs-jrnls,literature}

\begin{thebibliography}{10}
\providecommand{\url}[1]{#1}
\csname url@samestyle\endcsname
\providecommand{\newblock}{\relax}
\providecommand{\bibinfo}[2]{#2}
\providecommand{\BIBentrySTDinterwordspacing}{\spaceskip=0pt\relax}
\providecommand{\BIBentryALTinterwordstretchfactor}{4}
\providecommand{\BIBentryALTinterwordspacing}{\spaceskip=\fontdimen2\font plus
\BIBentryALTinterwordstretchfactor\fontdimen3\font minus
  \fontdimen4\font\relax}
\providecommand{\BIBforeignlanguage}[2]{{%
\expandafter\ifx\csname l@#1\endcsname\relax
\typeout{** WARNING: IEEEtran.bst: No hyphenation pattern has been}%
\typeout{** loaded for the language `#1'. Using the pattern for}%
\typeout{** the default language instead.}%
\else
\language=\csname l@#1\endcsname
\fi
#2}}
\providecommand{\BIBdecl}{\relax}
\BIBdecl

\bibitem{smith_staircase_2012}
B.~P. Smith, A.~Farhood, A.~Hunt, F.~R. Kschischang, and J.~Lodge, ``Staircase
  {{Codes}}: {{FEC}} for 100 {{Gb}}/s {{OTN}},'' \emph{J. Lightw. Technol.},
  vol.~30, no.~1, pp. 110--117, Jan. 2012.

\bibitem{richardson_design_2018-1}
T.~Richardson and S.~Kudekar, ``Design of {{Low}}-{{Density Parity Check
  Codes}} for {{5G New Radio}},'' \emph{{IEEE} Commun. Mag.}, vol.~56, no.~3,
  pp. 28--34, Mar. 2018.

\bibitem{gallagerPhD}
R.~G. Gallager, ``Low-density parity-check codes,'' \emph{IRE Trans. Inf.
  Theory}, vol.~8, no.~1, pp. 21--28, 1962.

\bibitem{richardson_capacity}
T.~J. Richardson and R.~L. Urbanke, ``The capacity of low-density parity-check
  codes under message-passing decoding,'' \emph{{IEEE} Trans. Inf. Theory},
  vol.~47, no.~2, pp. 599--618, 2001.

\bibitem{lechner_analysis_2012}
G.~Lechner, T.~Pedersen, and G.~Kramer, ``Analysis and {{Design}} of {{Binary
  Message Passing Decoders}},'' \emph{{IEEE} Trans. Commun.}, vol.~60, no.~3,
  pp. 601--607, Mar. 2012.

\bibitem{planjery2013finite}
S.~K. Planjery, D.~Declercq, L.~Danjean, and B.~Vasic, ``Finite alphabet
  iterative decoders -- part {I}: Decoding beyond belief propagation on the
  binary symmetric channel,'' \emph{{IEEE} Trans. Commun.}, vol.~61, no.~10,
  pp. 4033--4045, Oct. 2013.

\bibitem{Bauch}
J.~{Lewandowsky} and G.~{Bauch}, ``{Information-Optimum LDPC Decoders Based on
  the Information Bottleneck Method},'' \emph{IEEE Access}, vol.~6, pp.
  4054--4071, Jan. 2018.

\bibitem{benyacoubTMP}
E.~{Ben Yacoub}, F.~Steiner, B.~Matuz, and G.~Liva, ``Protograph-based {LDPC}
  code design for ternary message passing decoding,'' in \emph{Proc. ITG Int.
  Conf. Syst. Commun. Coding (SCC)}, Rostock, Germany, Feb. 2019.

\bibitem{steiner_jlt19}
F.~{Steiner}, E.~{Ben Yacoub}, B.~{Matuz}, G.~{Liva}, and A.~G. i.~{Amat},
  ``One and two bit message passing for {SC-LDPC} codes with higher-order
  modulation,'' \emph{J. Lightw. Technol.}, vol.~37, no.~23, pp. 5914--5925,
  Dec 2019.

\bibitem{Hagenauer96}
J.~{Hagenauer}, E.~{Offer}, and L.~{Papke}, ``Iterative decoding of binary
  block and convolutional codes,'' \emph{{IEEE} Trans. Inf. Theory}, vol.~42,
  no.~2, pp. 429--445, Mar. 1996.

\bibitem{Fossorier02minsum}
{Jinghu Chen} and M.~P.~C. {Fossorier}, ``Near optimum universal belief
  propagation based decoding of low-density parity check codes,'' \emph{{IEEE}
  Trans. Commun.}, vol.~50, no.~3, pp. 406--414, Aug. 2002.

\bibitem{jones_approximate-min_2003}
C.~Jones, E.~Valles, M.~Smith, and J.~Villasenor, ``Approximate-{{MIN}}
  constraint node updating for {{LDPC}} code decoding,'' in \emph{Proc. IEEE
  Military Commun. Conf.}, Boston, MA, USA, Oct. 2003.

\bibitem{Lechner06minsum}
G.~{Lechner} and J.~{Sayir}, ``Improved sum-min decoding for irregular {LDPC}
  codes,'' in \emph{Proc. 4th International Symposium on Turbo Codes Related
  Topics}, Munich, Germany, 2006.

\bibitem{ryan_channel_codes}
W.~Ryan and S.~Lin, \emph{Channel Codes: Classical and Modern}.\hskip 1em plus
  0.5em minus 0.4em\relax {Cambridge University Press}, 2009.

\bibitem{kschischang_minsum}
B.~{Smith}, F.~R. {Kschischang}, and {Wei Yu}, ``Low-density parity-check codes
  for discretized min-sum decoding,'' in \emph{23rd Biennial Symp. Commun.},
  May 2006, pp. 14--17.

\bibitem{ashikhmin_EXIT}
A.~Ashikhmin, G.~Kramer, and S.~{ten Brink}, ``Extrinsic information transfer
  functions: Model and erasure channel properties,'' \emph{{IEEE} Trans. Inf.
  Theory}, vol.~50, no.~11, pp. 2657--2673, 2004.

\bibitem{Kudekar_saturated}
S.~{Kudekar}, T.~{Richardson}, and A.~R. {Iyengar}, ``Analysis of saturated
  belief propagation decoding of low-density parity-check codes,'' \emph{{IEEE}
  Trans. Inf. Theory}, vol.~63, no.~9, pp. 5734--5751, Jul. 2017.

\bibitem{Hu_peg}
X.-Y. Hu, E.~{Eleftheriou}, and D.~. {Arnold}, ``Progressive edge-growth tanner
  graphs,'' in \emph{Proc. IEEE Global Telecommun. Conf. (GLOBECOM)}, vol.~2,
  Nov 2001, pp. 995--1001 vol.2.

\bibitem{polyanskiy_channel_2010}
Y.~Polyanskiy, H.~V. Poor, and S.~Verdu, ``Channel {{Coding Rate}} in the
  {{Finite Blocklength Regime}},'' \emph{{IEEE} Trans. Inf. Theory}, vol.~56,
  no.~5, pp. 2307--2359, May 2010.

\end{thebibliography}

\end{document}